\documentstyle[12pt]{article}
\begin{document}

\begin{flushright}
\baselineskip=12pt
ACT-05/96\\
DOE/ER/40717--26\\
CTP-TAMU-13/96\\
\tt hep-ph/9605419
\end{flushright}

\begin{center}
\vglue 1cm
{\Large\bf Enhanced Supersymmetric Corrections to Top-Quark Production at the
Tevatron\\}
\vglue 1.cm
{\large Jaewan Kim$^1$, Jorge L. Lopez$^2$, D.V. Nanopoulos$^{1,3}$,\\ and
Raghavan Rangarajan$^1$}
\vglue 1cm
\begin{flushleft}
$^1$Astroparticle Physics Group, Houston Advanced Research Center (HARC)\\
The Mitchell Campus, The Woodlands, TX 77381, USA\\
$^2$Department of Physics, Bonner Nuclear Lab, Rice University\\ 6100 Main
Street, Houston, TX 77005, USA\\
$^3$Center for Theoretical Physics, Department of Physics, Texas A\&M
University\\ College Station, TX 77843--4242, USA\\
\end{flushleft}
\end{center}

\vglue .75cm
\begin{abstract}
We calculate the one-loop supersymmetric electroweak-like and QCD-like corrections to the top-quark pair-production cross section at the Tevatron, including the important effects of non-degenerate top-squarks and left-right top-squark mixing. The largest electroweak-like effects yield a negative shift in the cross section and are enhanced right below the threshold for top-quark decay into top-squark and higgsino-like neutralino, and can be as large as $-35\%$. The largest QCD-like effects are positive and are enhanced for light top-squark masses, and can be as large as $20\%$. Such shifts greatly exceed the present theoretical uncertainty in the Standard Model prediction, and therefore may be experimentally observable. We also explore the one-loop shifts in scenarios containing light top-squarks and higgsino-like neutralinos that have been recently proposed to explain various apparent experimental anomalies.
\end{abstract}
\vspace{0.5cm}

\begin{flushleft}
E-mail addresses:\\
\small\tt
\baselineskip=12pt
jaewan@diana.tdl.harc.edu\\
lopez@physics.rice.edu\\
dimitri@phys.tamu.edu\\
raghu@diana.tdl.harc.edu
\end{flushleft}

\newpage
\setcounter{page}{1}
\pagestyle{plain}
\baselineskip=14pt

\section{Introduction}
\label{sec:introduction}
The existence of the top quark has now been firmly established by the CDF and
D0 Collaborations at Fermilab \cite{CDFD0Top}. The analysis of over $100\,{\rm
pb}^{-1}$ of data by each experiment has allowed a determination of the
top-quark mass and pair-production cross section \cite{CDFD0LaThuile}
\begin{eqnarray}
m^{\rm CDF}_t&=&176\pm\hphantom{1}9\,{\rm GeV}\,, \qquad
\sigma^{\rm CDF}_{t\bar t}=7.6^{+1.9}_{-1.5}\hphantom{1.}\,{\rm pb}\,;
\label{eq:CDF}\\
m^{\rm D0\hphantom{F}}_t&=&170\pm18\,{\rm GeV}\,, \qquad
\sigma^{\rm D0\hphantom{i1}}_{t\bar t}=5.2\pm1.8\,{\rm pb}\,;
\label{eq:D0}
\end{eqnarray}
to better than $\sim5\%$ and $\sim20\%$ respectively. The experimental error
flags are expected to be further reduced when the data set is fully analyzed. A
significant increase in sensitivity will become available once the Main
Injector upgrade becomes operational in 1999, with an expected reduction in the
top-quark mass uncertainty down to $3.5\,(2.0)\,{\rm GeV}$ and in the
production cross section down to $11\%\,(6\%)$ with an integrated luminosity of
${\cal L}=1\,(10)\,{\rm fb}^{-1}$ \cite{TeV33}. On the other hand, the
theoretical prediction for the cross section in the Standard Model has become
rather precise \cite{Laenen,Berger,Catani}, and is presently known to better
than 10\% for fixed values of $m_t$, {\em e.g.} \cite{Berger}
\begin{equation}
\sigma_{t\bar t}^{\rm theory}\,[170]=6.48^{+0.09}_{-0.48}\,\,{\rm pb}\,,\qquad
\sigma_{t\bar t}^{\rm theory}\,[175]=5.52^{+0.07}_{-0.42}\,\,{\rm pb}\,.
\label{eq:sigma}
\end{equation}
The agreement between theoretical expectations and experimental observations
is not tight enough to preclude moderate shifts ($\sim2\,{\rm pb}$)
to the production cross section from new physics phenomena. However, greater
than $50\%$ shifts in the Standard Model prediction due to new physics
may conflict with experimental observations. It is then opportune to investigate how large a shift scenarios for new physics may yield, most interestingly in the case of low-energy supersymmetry.

In this paper we consider the one-loop corrections to the pair-production cross
section for top quarks at the Tevatron, including both supersymmetric
electroweak-like (loops of charginos and bottom-squarks or neutralinos and top-squarks) and QCD-like (loops of
gluinos and squarks) contributions. Our calculation extends and corrects those
in the literature \cite{YangLi,SUSYQCD}, identifies the largest contributions
and determines under what conditions they may be enhanced, and treats all
important and realistic effects (such as top-squark mass non-degeneracy and
mixing) in a unified way. We also consider specific scenarios with especially
light supersymmetric particles that have been proposed in connection with the
$R_b$ observable \cite{Rbsusy}, an as-yet unexcluded light higgsino window
\cite{Feng}, and a possible supersymmetric explanation \cite{Kane,KaneMrenna}
for the $ee\gamma\gamma$ event observed by CDF \cite{Park}.
We stress that light supersymmetric particles that affect the top-quark cross
section are also likely to affect its decay width via new supersymmetric
channels, and the resulting decrease in $B(t\to bW)$ may affect the yield of
$bWbW$ events in a much more significant way than a shift in the underlying
production cross section.

This paper is organized as follows. In Sec.~\ref{sec:analytical} we present
compact expressions for the results of our analytical calculations for the
one-loop supersymmetric electroweak-like and QCD-like corrections and contrast them
with existing calculations. In Sec.~\ref{sec:numerical} we present a
qualitative and quantitative study of the expected corrections and identify
regions in parameter space that entail the largest possible effects. These effects may easily exceed the theoretical uncertainty in the Standard Model prediction for the top-quark cross section, and thus may be disentangled at future Tevatron runs. In Sec.~\ref{sec:models} we adapt our general results to study specific models that have been proposed in the literature and that contain especially light supersymmetric particles. Sec.~\ref{sec:conclusions}
summarizes our conclusions and the Appendix contains some further expressions.

\section{Analytical results}
\label{sec:analytical}
\begin{figure}[p]
\vspace{5in}
\includegraphics{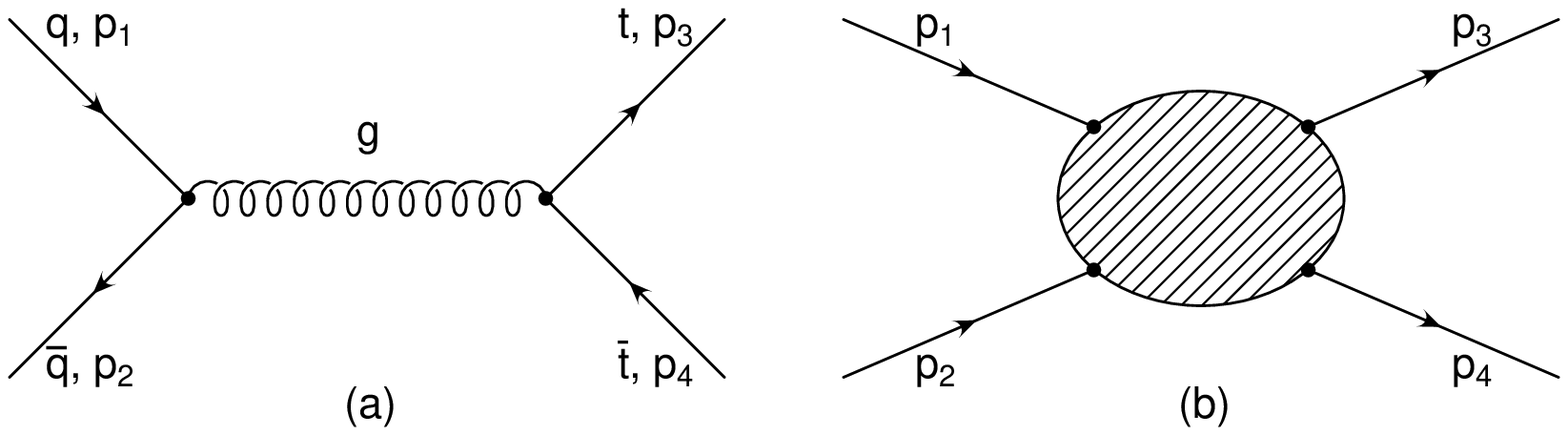}
\caption{Tree-level diagram (a) describing $q\bar q\to t\bar t$ production. The
blob diagram (b) indicates the choice of external momenta used throughout our calculations.}
\label{fig:tree}
\end{figure}

\begin{figure}[p]
\vspace{6in}
\includegraphics{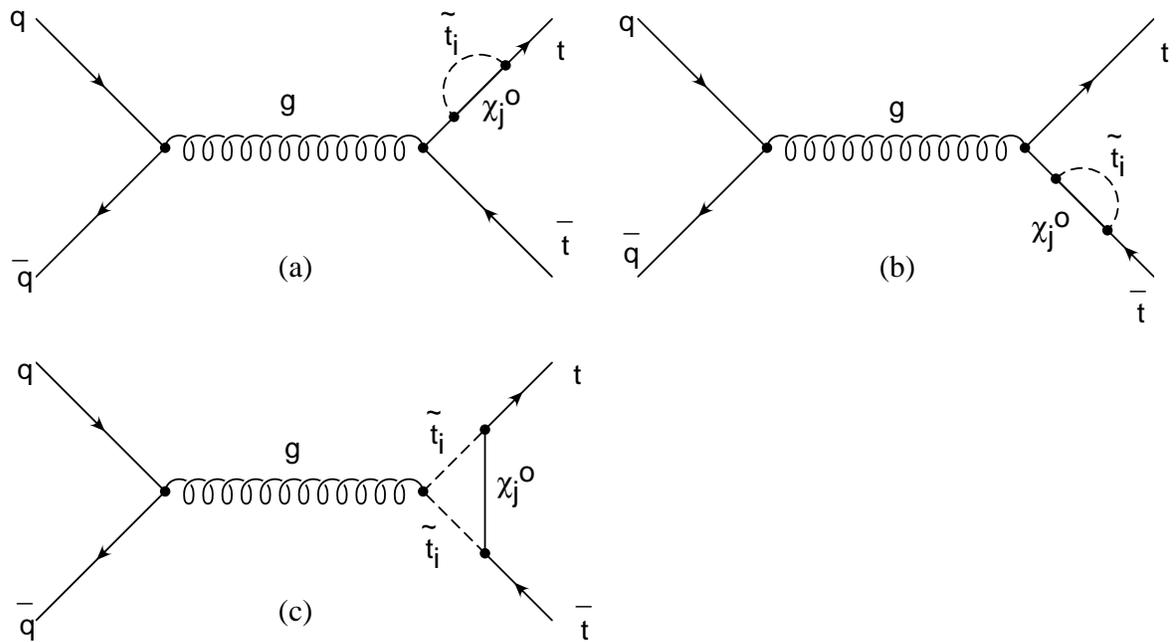}
\caption{Feynman diagrams describing the one-loop supersymmetric electroweak-like corrections to $q\bar q\to t\bar t$, including external leg corrections (a,b) and vertex corrections (c) from top-squark and neutralino
loops. An analogous set of diagrams exists where the top-squarks are replaced
by bottom-squarks and the neutralinos by charginos.}
\label{fig:EWdiagrams}
\end{figure}

\begin{figure}[p]
\vspace{6in}
\includegraphics{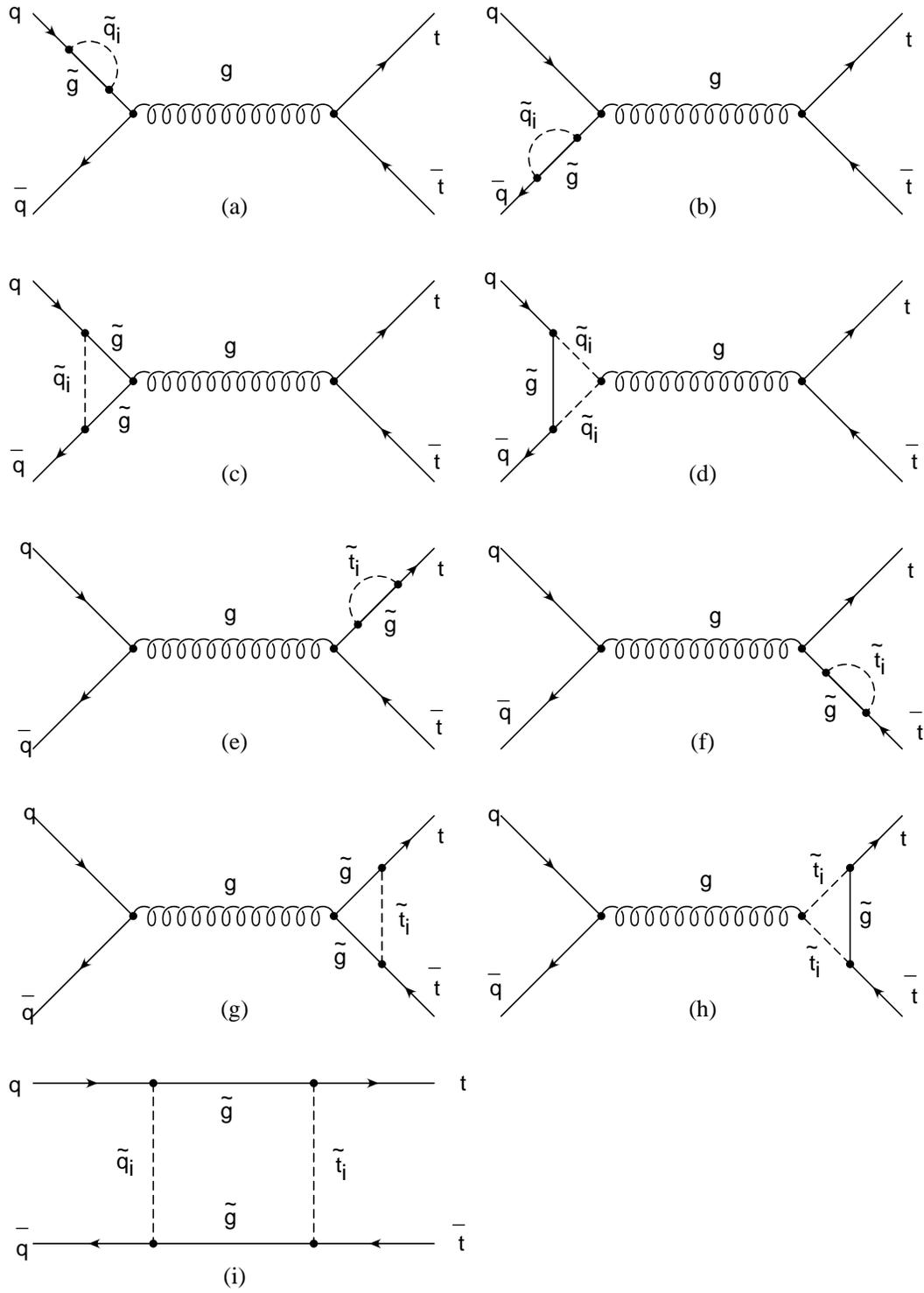}
\vspace{4cm}
\caption{Feynman diagrams describing the one-loop supersymmetric QCD-like corrections to $q\bar q\to t\bar t$, including external leg corrections (a,b,e,f), vertex corrections (c,d,g,h), and a representative box diagram (i).}
\label{fig:QCDdiagrams}
\end{figure}

Top-quark pair-production at a hadron collider proceeds at tree-level via two
underlying $s$-channel parton processes: $q\bar q$ annihilation (see Fig.~\ref{fig:tree}(a)) and gluon fusion. At the Tevatron the first process dominates ($90\%$) \cite{Laenen}, and therefore we
neglect the latter in what follows. One-loop corrections to $q\bar q\to g\to
t\bar t$ in the context of low-energy supersymmetry fall into two categories:
electroweak-like \cite{YangLi} and QCD-like \cite{SUSYQCD}. The significant
electroweak-like vertex and external leg corrections modify the outgoing part of the Feynman diagram via new contributions to the $gt\bar t$ vertex and to the normalization of the top-quark wavefunction that involve loops of top-squarks and neutralinos or bottom-squarks and charginos (see Fig.~\ref{fig:EWdiagrams}). The QCD-like corrections modify both the incoming and outgoing portions of the diagram in an analogous fashion, but involve
loops of gluinos and first-generation-squarks or gluinos and top-squarks (see Fig.~\ref{fig:QCDdiagrams}). There are also (smaller) QCD-like box diagram contributions involving gluinos and squarks (see Fig.~\ref{fig:QCDdiagrams}(i)).

The tree-level $q\bar q\to t\bar t$ parton cross section is given by
\begin{equation}
\widehat\sigma={8\pi\alpha^2_s\over9\widehat s^2}\,\beta_t\,
{\textstyle{1\over3}}(\widehat s+2m^2_t)
\label{eq:tree}
\end{equation}
where $\beta_t=\sqrt{1-4m^2_t/\widehat s}$, and $\sqrt{\widehat s}$ is the
usual parton-level center-of-mass energy. The supersymmetric electroweak-like
one-loop corrections involving the Higgsino components of the neutralinos and
charginos are enhanced by a large top-quark Yukawa coupling as seen below
\begin{equation}
\lambda_t\, 
\left(\!\!\!\begin{array}{c}\widehat t_L\\ \widehat b_L\end{array}\!\!\!\right)
\widehat H_2\, \widehat t_R\ \ni\ \lambda_t\, \widetilde b_L \widetilde
H^\pm\, t_R\,,\ \lambda_t\, t_L \widetilde H^0_2\,\widetilde t_R\,,\
\lambda_t\,\widetilde t_L \widetilde H^0_2\, t_R\ ,
\label{eq:Yukawa}
\end{equation}
where the carets represent superfields and the terms on the right-hand side
include the Yukawa coupling components of interest.
These corrections involve loops of top-squarks and neutralinos or bottom-squarks and charginos and are given by
\begin{equation}
\Delta\widehat\sigma^{\rm EW}={8\pi\alpha^2_s\over9\widehat s^2}\,\beta_t\,
\left({\lambda_t\over4\pi}\right)^2
\left[{\textstyle{2\over3}}(\widehat s+2m^2_t)(F_1^n+F_1^c)
+2(F_5^n+F_5^c) m_t\widehat s\right]\ ,
\label{eq:EWloop}
\end{equation}
where $\lambda_t=g m_t/(\sqrt{2}M_W\sin\beta)$ is the top-quark Yukawa
coupling, and $\tan\beta=v_2/v_1$ where $v_1,v_2$ are the Higgs vacuum expectation values that arise in the MSSM. Also,  $F_{1,5}^{n,c}$ are form factors that encode the loop functions and depend on the various mass parameters. The top-squark--neutralino form factors are given by:
\begin{eqnarray}
F_1^n&=&\sum_{j=1}^4\biggl\{ N_{j4}N^*_{j4}\sum_{J=1}^2
\left[ c_{24}+m^2_t (c_{11}+c_{21})+{\textstyle{1\over2}}B_1
+m^2_t B'_1\right]^{(\chi^0_j,\tilde t_J,\tilde t_J)}\nonumber\\
&&-{\rm Re}\,(N_{j4}N_{j4})\sum_{J=1}^2
(-1)^{J+1}\sin(2\theta_t)\, m_t\, m_{\chi^0_j}
\,(c_0+c_{11}+B'_0)^{(\chi^0_j,\tilde t_J,\tilde t_J)}\biggr\}\ ;
\label{eq:F1n}\\
F_5^n&=&\sum_{j=1}^4\biggl\{ N_{j4}N^*_{j4}\sum_{J=1}^2
\left[-{\textstyle{1\over2}}m_t(c_{11}+c_{21})\right]
^{(\chi^0_j,\tilde t_J,\tilde t_J)}\nonumber\\
&&+{\rm Re}\,(N_{j4}N_{j4})\sum_{J=1}^2
(-1)^{J+1}\,{\textstyle{1\over2}}
\sin(2\theta_t)\,m_{\chi^0_j}\,
(c_0+c_{11})^{(\chi^0_j,\tilde t_J,\tilde t_J)}\biggr\}\ .
\label{eq:F5n}
\end{eqnarray}
Here $N_{j4}$ represents the higgsino admixture of $\chi^0_j$ ($\widetilde
H^0_2$ in the notation of Eq.~(\ref{eq:Yukawa}) and Ref.~\cite{HK}), and the
$J=1,2$ sum runs over the two top-squark mass eigenstates ($m_{\tilde
t_{1,2}}$), which
are obtained from the $\tilde t_{L,R}$ gauge eigenstates via: $\tilde
t_1=\cos\theta_t\,\tilde t_L+\sin\theta_t\,\tilde t_R$ and $\tilde
t_2=-\sin\theta_t\,\tilde t_L+\cos\theta_t\,\tilde t_R$.
The various $B$ and $c$ functions in the above expressions are the well
documented Passarino-Veltman functions \cite{PV} (adapted to our metric where
$p^2_i=m^2_i$); the $B$ functions depend on $(m_t,m_{\chi^0_j},m_{\tilde t_J})$
whereas the $c$ functions depend on $(-p_3,p_3+p_4,m_{\chi^0_j},
m_{\tilde t_J},m_{\tilde t_J})$ [as reminded by the superscripts in
Eqs.~(\ref{eq:F1n},\ref{eq:F5n})], where $p_3$ and $p_4$ are the momenta of the
outgoing top-quark and anti-top-quark respectively (see Fig.~\ref{fig:tree}(b)).
The bottom-squark--chargino form factors are in turn given by
\begin{eqnarray}
F_1^c&=&\sum_{j=1}^2 V_{j2}V^*_{j2}\sum_{J=1}^2
\left(\begin{array}{c}\cos^2\theta_b^{\ (J=1)}\\
\sin^2\theta_b^{\ (J=2)}\end{array}\right)
\left[c_{24}+m^2_t(c_{11}+c_{21})+{\textstyle{1\over2}}B_1+m^2_t B'_1
\right]^{(\chi^\pm_j,\tilde b_J,\tilde b_J)}
\label{eq:F1c}\\
F_5^c&=&\sum_{j=1}^2 V_{j2}V^*_{j2}\sum_{J=1}^2
\left(\begin{array}{c}\cos^2\theta_b^{\ (J=1)}\\
\sin^2\theta_b^{\ (J=2)}\end{array}\right)
\left[-{\textstyle{1\over2}}m_t(c_{11}+c_{21})
\right]^{(\chi^\pm_j,\tilde b_J,\tilde b_J)}\ ,
\label{eq:F5c}
\end{eqnarray}
where $V_{j2}$ represents the higgsino admixture of $\chi^\pm_j$, and for
completeness we have allowed a non-vanishing bottom-squark mixing angle
such that $\tilde b_1=\cos\theta_b\,\tilde b_L+\sin\theta_b\,\tilde b_R$ and
$\tilde b_2=-\sin\theta_b\,\tilde b_L+\cos\theta_b\,\tilde b_R$. The $J=1,2$
sum runs over the two bottom-squark mass eigenstates ($m_{\tilde b_{1,2}}$).
Note that if $\theta_b=0$, only the left-handed bottom-squark is involved in
the loops. In Eqs.~(\ref{eq:F1c},\ref{eq:F5c}) the $B$ functions depend on $(m_t,m_{\chi^\pm_j},m_{\tilde b_J})$ whereas the $c$ functions depend on $(-p_3,p_3+p_4,m_{\chi^\pm_j},m_{\tilde b_J},m_{\tilde b_J})$. 

The functions $B_1$ and $c_{24}$ in Eqs.~(\ref{eq:F1n},\ref{eq:F1c}) contain infinities. We used the modified minimal subtraction scheme and introduced counterterms to eliminate them. Nevertheless, it turns out that the infinities (and the renormalization-scale $\mu$ dependence) of these two functions cancel each other out in both equations (independently) even without introducing counterterms.

Since the Passarino-Veltman functions can be
notoriously difficult to evaluate numerically for certain values of the
parameters, we have employed different methods (all of which agree), in
particular using the software package {\tt ff} \cite{vanOld}. Our results
for the case of no top- or bottom-squark mixing ($\theta_t=\theta_b=0$)
disagree with those originally published in Ref.~\cite{YangLi}. However, these
authors have since revised their calculation and their corrected expressions
\cite{YangLiE} are now in agreement with our results above. Our results for the
realistic case of squark mixing ($\theta_t,\theta_b\not=0$) are
new.\footnote{While the present paper was being written up Ref.~\cite{YangLiLR}
appeared, which contains a calculation of the $\theta_t\not=0$ case, that
agrees with our result above in the case of real $N_{j4}$ values.}

In Eqs.~(\ref{eq:EWloop},\ref{eq:F1n},\ref{eq:F5n},\ref{eq:F1c},\ref{eq:F5c})
we have not explicitly exhibited the additional contributions to the
$F^{n,c}_{1,5}$ form factors that arise from the gaugino admixtures of the
neutralinos and charginos. These are proportional to the electroweak gauge
couplings and are therefore not enhanced by a large top-quark Yukawa coupling.
These contributions have been neglected in our present analysis. For completeness, in the Appendix we present analytic expressions for the contributions from all (higgsino and gaugino) admixtures to the vertex and external leg corrections. We have also not exhibited the electroweak-like corrections to the incoming part of the diagram, that involve loops of first-generation squarks and neutralinos or charginos. These corrections are rather small as the relation $m_q\approx m_{\tilde q}+m_\chi$, that the numerical analysis below shows is required for enhancement, is not satisfied in this case. Furthermore, for higgsino-like neutralinos these contributions are proportional to $m^2_q$ ({\em c.f.} Eq.~(\ref{eq:EWloop})) and are therefore negligible.

The QCD-like corrections involve self-energy, vertex, and box diagrams. In the
approximation of neglecting the box diagrams,\footnote{Recent calculations
of supersymmetric QCD-like corrections to light-quark scattering at the Tevatron
indicate that box diagram contributions are indeed small near the threshold
region \cite{ER}.} one can readily obtain the shifts in the total parton-level
cross sections and cast them in the same format as
the electroweak-like corrections  given above. In this form comparison
between the two classes of corrections becomes transparent. The QCD-like
one-loop correction becomes
\begin{equation}
\Delta\widehat\sigma^{\rm QCD}={8\pi\alpha^2_s\over9\widehat s^2}\,\beta_t\,
\left({\alpha_s\over4\pi}\right)
\left[{\textstyle{2\over3}}(\widehat s+2m^2_t)(F^t_1+F^q_1)+2F^t_5 m_t
\widehat s\right]\ .
\label{eq:QCDloop}
\end{equation}
The form factor $F^t_1$, which modifies the outgoing part of the diagram is
given by
\begin{eqnarray}
F^t_1&=&\sum_{J=1}^2\Bigl\{(-{\textstyle{1\over3}})
[c_{24}+m^2_t(c_{11}+c_{21})]^{(\tilde g\tilde t_J\tilde t_J)}
+({\textstyle{4\over3}})[B_1+2m^2_t B'_1]
\nonumber\\
&&\qquad+(-1)^J\sin(2\theta_t)m_t m_{\tilde g}
[({\textstyle{1\over3}})(c_0+c_{11})^{(\tilde g\tilde t_J\tilde t_J)}
-({\textstyle{8\over3}})B'_0]\nonumber\\
&&+({\textstyle{3\over2}})[-{\textstyle{1\over2}}+2c_{24}
+\widehat s(c_{22}-c_{23})
-m^2_{\tilde g}c_0-m^2_t(c_0+2c_{11}+c_{21})\nonumber\\
&&\qquad-(-1)^J\,2\sin(2\theta_t)m_t m_{\tilde g}(c_0+c_{11})
]^{(\tilde t_J\tilde g\tilde g)}
\Bigr\}\ ,
\label{eq:Ft1}
\end{eqnarray}
where $\theta_t$ is the top-squark mixing angle, the $B$ functions depend on
$(m_t,m_{\tilde g},m_{\tilde t_J})$, and the superscripts $(abb)$
indicate that the corresponding $c$ functions depend on
$(-p_3,p_3+p_4,m_a,m_b,m_b)$. 

The analogous form factor that modifies the incoming part of the diagram
($F^q_1$) is obtained from $F^t_1$ by setting $\tilde t_J\to\tilde q_J$,
$m_t\to m_q$, $\theta_t\to\theta_q$, and by replacing $p_3,p_4$ by
$p_1,p_2$, the incoming quark and anti-quark momenta. Since $m_q\approx0$,
the expression for $F^q_1$ simplifies considerably. Note the close resemblance
between the first two lines in $F_1^t$ and the expression for $F^n_1$ in
Eq.~(\ref{eq:F1n}) above, as both form factors originate from analogous
diagrams ($\chi^0_j\leftrightarrow\tilde g$); the differences in the
coefficients stem from the color factors that appear in the QCD-like
diagram. In contrast, the last two lines in $F_1^t$ correspond to a new diagram (not present in the electroweak-like case) where the gluino couples directly to the gluon and which carries a large color factor (Fig.~\ref{fig:QCDdiagrams}(g)). We also have
\begin{eqnarray}
F^t_5&=&\sum_{J=1}^2\Bigl\{
({\textstyle{1\over6}})
[m_t(c_{11}+c_{21})-(-1)^J\sin(2\theta_t)m_{\tilde g}(c_0+c_{11})]
^{(\tilde g\tilde t_J\tilde t_J)}\nonumber\\
&&+({\textstyle{3\over2}})
[m_t(c_{11}+c_{21})+(-1)^J\sin(2\theta_t)m_{\tilde g}c_{11}]
^{(\tilde t_J\tilde g\tilde g)}\Bigr\}\ ,
\label{eq:F5}
\end{eqnarray}
where we again note the resemblance (up to the color factor coefficient)
between the first term in $F^t_5$ and its counterpart $F^n_5$ in
Eq.~(\ref{eq:F5n}). The second term in $F^t_5$ corresponds to the diagram not
present in the electroweak-like case. The analogous form factor $F^q_5$ (that
arises from the incoming part of the diagram) is obtained from $F^t_5$ by
setting $\tilde t_J\to\tilde q_J$, $m_t\to m_q$, $\theta_t\to\theta_q$, by
introducing an overall minus sign and by replacing $p_3,p_4$ by $p_1,p_2$, the
incoming quark and anti-quark momenta. This contribution is negligible since
$m_q\approx0$ and because in the MSSM $\theta_q\propto m_q\approx0$. Our
results for the supersymmetric QCD-like corrections agree with those presented
earlier in  Ref.~\cite{SUSYQCD}, and its erratum \cite{SUSYQCDe}. 

\begin{figure}[p]
\vspace{5in}
\includegraphics{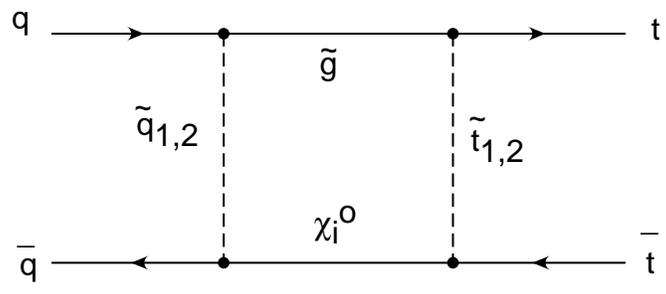}
\caption{Representative box diagram (one of four possible ones) describing the mixed one-loop supersymmetric QCD-like and electroweak-like corrections.}
\label{fig:EWQCD}
\end{figure}

For general neutralino composition there exist box diagrams that mix electroweak-like and QCD-like corrections,\footnote{We thank C. Kao for point this out to us.} with gluinos, squarks, top-squarks, and neutralinos in the loop (see Fig.~\ref{fig:EWQCD}). For the higgsino-like neutralino case that we consider, these diagrams are proportional to the light-quark Yukawa coupling and therefore negligible.

The actual observable cross section is obtained by integrating
$\widehat\sigma+\Delta\widehat\sigma$ over the parton distribution functions,
{\em i.e.},
\begin{eqnarray}
\sigma+\Delta\sigma&=&\int_{\tau_0}^1 d\tau\int_\tau^1 {dx_1\over x_1}
\Bigl[
u(x_1)u(x_2)+u(x_1)sea(x_2)+sea(x_1)u(x_2)\nonumber\\
&&+d(x_1)d(x_2)+d(x_1)sea(x_2)+sea(x_1)d(x_2)\nonumber\\
&&+6sea(x_1)sea(x_2)\Bigr](\widehat\sigma+\Delta\widehat\sigma)(\widehat s)\ ,
\label{eq:PDF}
\end{eqnarray}
where $\tau_0=4m^2_t/s$, $x_2=\tau/x_1$, $\widehat s=\tau s$, and the parton
distribution functions ($u,d,sea$) are taken from Ref.~\cite{MT} setting the
scale $Q=m_t$.

\section{Numerical results}
\label{sec:numerical}
Inspecting the above formulas, one can immediately get an idea of the typical
size of the one-loop corrections, as they are proportional to
\begin{equation}
{\Delta\widehat\sigma\over\widehat\sigma}^{\rm EW}\propto
\left({\lambda_t\over4\pi}\right)^2\,,\qquad
{\Delta\widehat\sigma\over\widehat\sigma}^{\rm QCD}\propto
\left({\alpha_s\over4\pi}\right)\ .
\label{eq:typical}
\end{equation}
For the favored values of $m_t$, both these factors are $\sim1\%$. We thus
see that unless there are enhancements within the loop factors
contributing to these diagrams, both contributions are comparable in
size, and more importantly, much too small to be disentangled from the
Standard Model contribution, or to be observed experimentally at the Tevatron
or any of its planned or proposed upgrades.

Dynamical enhancements are however possible for restricted ranges
of the mass parameters. The top-quark self-energy diagrams show one such
enhancement through the $B'_0,B'_1$ functions when $m_t\approx m_a+m_b$, where
$m_{a,b}$ are the masses of the two particles entering the two-point function.
The enhancement occurs right below the threshold for $t\to a+b$ decay. For the
top quark there are three possibilities for $m_{a,b}$:
$(m_{\tilde t_{1,2}},m_{\chi^0_{1,2,3,4}})$;
$(m_{\tilde b_{1,2}},m_{\chi^\pm_{1,2}})$; and
$(m_{\tilde t_{1,2}},m_{\tilde g})$.
Given the present experimental lower limits on the squark (excluding $\tilde
t$) and gluino masses ({\em i.e.}, $m_{\tilde q},m_{\tilde g}>175\,{\rm GeV}$;
$m_{\tilde q}\approx m_{\tilde g}>230\,{\rm GeV}$ \cite{TeVsqg}), only the
first possibility may be realized. That is, enhancements may occur for
$m_t\approx m_{\tilde t_{1,2}}+m_{\chi^0_{1,2,3,4}}$. Moreover, these
enhancements will be maximized when the neutralinos have a high higgsino
content. (We do not see such enhancements for the incoming part of the diagram
because $m_q\approx0$.)

In this region of parameter space (and for real values of $N_{j4}$) one obtains
the following approximate expression
\begin{eqnarray}
{\Delta\widehat\sigma\over\widehat\sigma}^{\rm EW}
&\approx&\left({\lambda_t\over4\pi}\right)^2
\sum_{j=1}^4(N_{j4})^2\sum_{J=1}^2 2\left[m^2_t B'_1
-(-1)^{J+1}\sin(2\theta_t)m_tm_{\chi^0_j}B'_0\right]\nonumber\\
&\to&\left({\lambda_t\over4\pi}\right)^2\sum_{J=1}^2 2\left[m^2_t B'_1
-(-1)^{J+1}\sin(2\theta_t)m_tm_\chi B'_0\right]\ ,
\label{eq:ratio}
\end{eqnarray}
where the second expression follows when one of the neutralinos ($\chi$)
carries the full higgsino admixture. Also, the choice $\tan\beta=1$ maximizes
$\lambda_t$ for a fixed value of $m_t$. In this `best case scenario', one can
plot  $\Delta\sigma/\sigma$ ({\em i.e.}, after integration over parton
distribution functions) versus the neutralino mass ($m_\chi$), and study the
dependence on the top-squark masses and mixing angle. Note that the
mixing-angle term does not contribute if the top-squark masses are degenerate.

\begin{figure}[p]
\vspace{6in}
\includegraphics{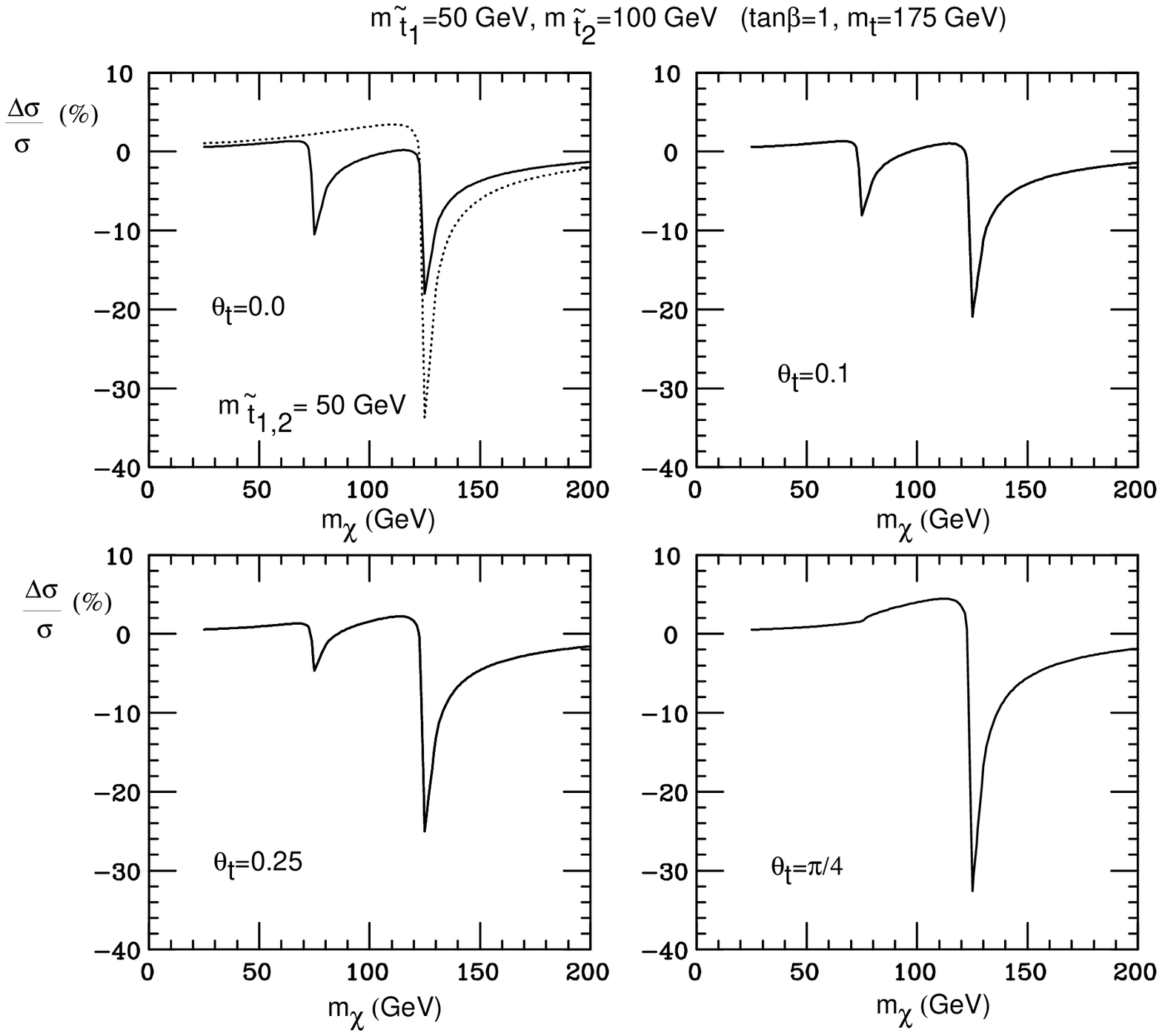}
\caption{The relative (\%) one-loop supersymmetric electroweak-like correction to
the top-quark pair-production cross section at the Tevatron as a function of
the (higgsino-like) neutralino mass, for $m_t=175\,{\rm GeV}$,
$\tan\beta=1$, $m_{\tilde t_1}=50\,{\rm GeV}$, $m_{\tilde t_2}=100\,{\rm GeV}$,
and various choices of the top-squark mixing angle
($\theta_t=0.0,0.10,0.25,{\pi\over4}$). The dotted curve on the upper-left-hand
panel corresponds to $m_{\tilde t_1}=m_{\tilde t_2}=50\,{\rm GeV}$. Note the
dips on the curves when $m_t\approx m_{\tilde t_{1,2}}+m_\chi$.}
\label{fig:50100}
\end{figure}

\begin{figure}[p]
\vspace{6in}
\includegraphics{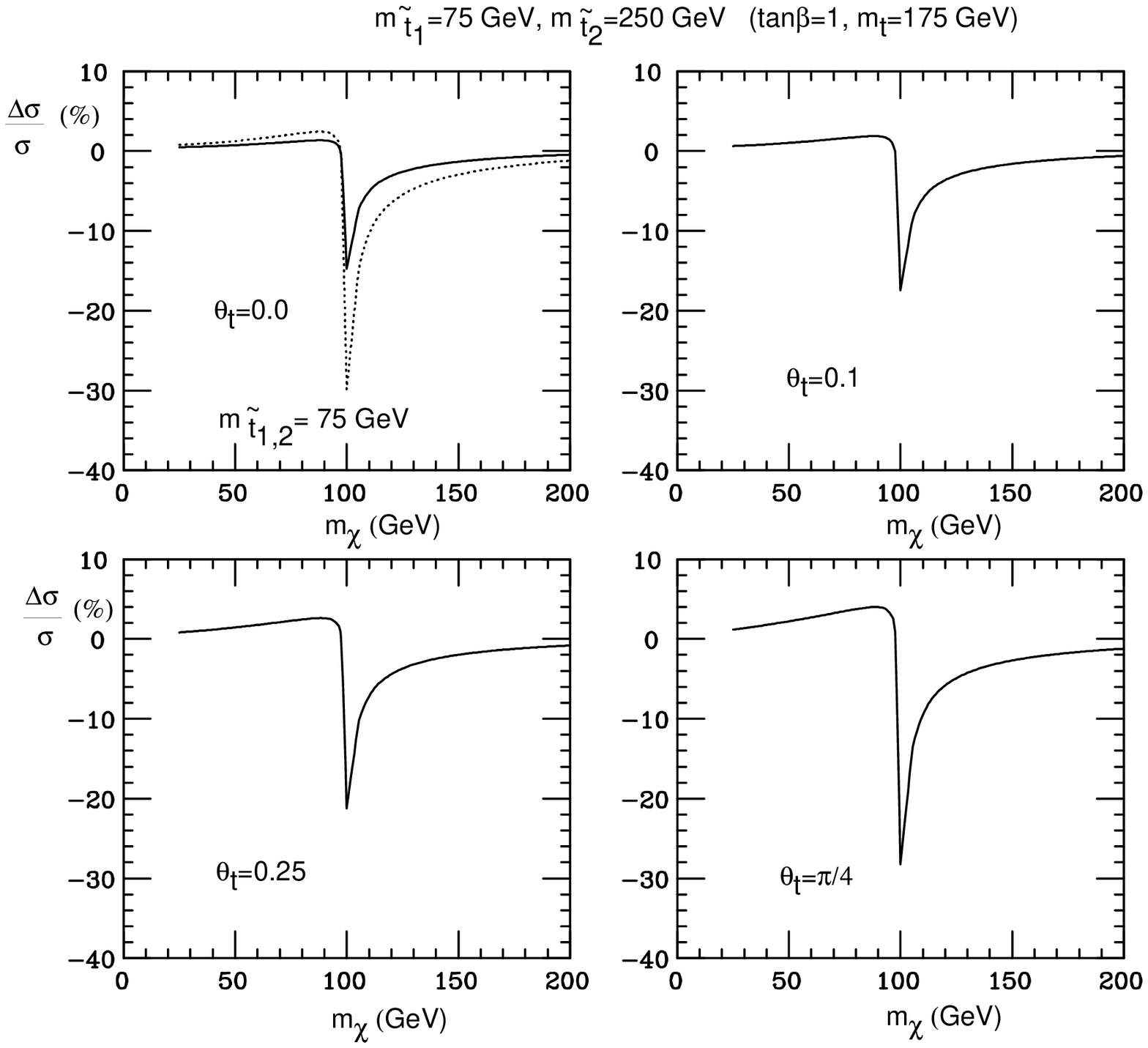}
\caption{The relative (\%) one-loop supersymmetric electroweak-like correction  to
the top-quark pair-production cross section at the Tevatron as a function of
the (higgsino-like) neutralino mass, for $m_t=175\,{\rm GeV}$,
$\tan\beta=1$, $m_{\tilde t_1}=75\,{\rm GeV}$, $m_{\tilde t_2}=250\,{\rm GeV}$,
and various choices of the top-squark mixing angle
($\theta_t=0.0,0.10,0.25,{\pi\over4}$). The dotted curve on the upper-left-hand
panel corresponds to $m_{\tilde t_1}=m_{\tilde t_2}=75\,{\rm GeV}$. Note the
dips on the curves when $m_t\approx m_{\tilde t_{1,2}}+m_\chi$.}
\label{fig:75250}
\end{figure}

In the case of degenerate masses, taken at
$m_{\tilde t_1}=m_{\tilde t_2}=50\,(75)\,{\rm GeV}$, the resulting relative
shift (\%), as a function of $m_\chi$, is shown by the dotted curve on the
upper-left-hand panel in Figs.~\ref{fig:50100}\,(\ref{fig:75250}).\footnote{The
numerical results in Figs.~\ref{fig:50100} and \ref{fig:75250} have been obtained for the `best case scenario' outlined below Eq.~(\ref{eq:ratio})
and include the complete electroweak-like vertex and external leg corrections.} The large dips at $m_\chi\approx
m_t-m_{\tilde t_{1,2}}\approx125\,(100)\,{\rm GeV}$
derive from the $B'_1$ term in Eq.~(\ref{eq:ratio}) and have been regularized
by setting $m^2_t\to m^2_t-im_t\Gamma_t$, where $\Gamma_t$ (a few GeV) is the
top-quark decay width.\footnote{Throughout our numerical calculations we have
used $m_t=175\,{\rm GeV}$ and $m_t\Gamma_t=289\,{\rm GeV}^2$. For $\alpha_s$ we took the world-average value of 0.118 at the scale $\mu=M_Z$ in the modified minimal subtraction scheme.} The depth of the dips depends on the value of $m_\chi/m_t$ at which it occurs, and is maximized for  $m_\chi/m_t\approx0.73$, which in this case implies $m_\chi\approx 128\,{\rm GeV}$, $m_{\tilde t}\approx 47\,{\rm GeV}$ ({\em i.e.}, as in Fig.~\ref{fig:50100} (dotted curve, upper-left-hand panel)).

In a more realistic scenario, light top-squarks cannot be degenerate in mass
because limits on additional contributions to the $\rho$ parameter restrict
the splitting between the $\tilde t_L$ and $\tilde b_L\approx \tilde b_1$, and
from direct experimenal searches one estimates that $m_{\tilde b_1}>200\,{\rm
GeV}$. In the case of no top-squark mixing ($\theta_t=0,{\pi\over2}$), the
splitting of $\tilde t_{1,2}$ leads to a double-dip structure if both
top-squark masses are such that $m_t\approx m_{\tilde t_{1,2}}+m_\chi$ can be
satisfied, as in Fig.~\ref{fig:50100} (solid curve, upper-left-hand panel),
where we have taken $m_{\tilde t_1}=50\,{\rm GeV}$ and $m_{\tilde
t_2}=100\,{\rm GeV}$. Note that the dips do not have the same depth, as
discussed above. In Fig.~\ref{fig:75250} there is a single dip because we have
taken $m_{\tilde t_1}=75\,{\rm GeV}$ and $m_{\tilde t_2}=250\,{\rm GeV}$.

Yet more realistic is the case of top-squark mixing, which is naturally present
in supergravity theories. This mixing tends to ``screen" the contributions of
top-squarks to the $Z$-pole observables \cite{Hollik}, and therefore allows
a larger $\tilde t_L-\tilde b_L$ mass splitting. In Figs.~\ref{fig:50100} and
\ref{fig:75250} we present the results for four choices of the mixing angle
$\theta_t=0,0.10,0.25$, and maximal mixing (${\pi\over4}$). The plots also
apply for $\theta_t\to{\pi\over2}-\theta_t$, which leaves $\sin(2\theta_t)$
unchanged. In this case the $B'_0$ function in Eq.~(\ref{eq:ratio}) plays an
important role, as it exhibits a similar dip behavior as $B'_1$ does, although
with a different sign at each dip. This effect makes one dip deeper
(corresponding to $\tilde t_1$) whereas the other one shallower (corresponding
to $\tilde t_2$), as evidenced in Fig.~\ref{fig:50100}. The effect can be very
significant, completely eliminating one of the dips at maximum mixing angle. In
Fig.~\ref{fig:75250} only the dip that gets deeper exists.

\begin{figure}[p]
\vspace{5in}
\includegraphics{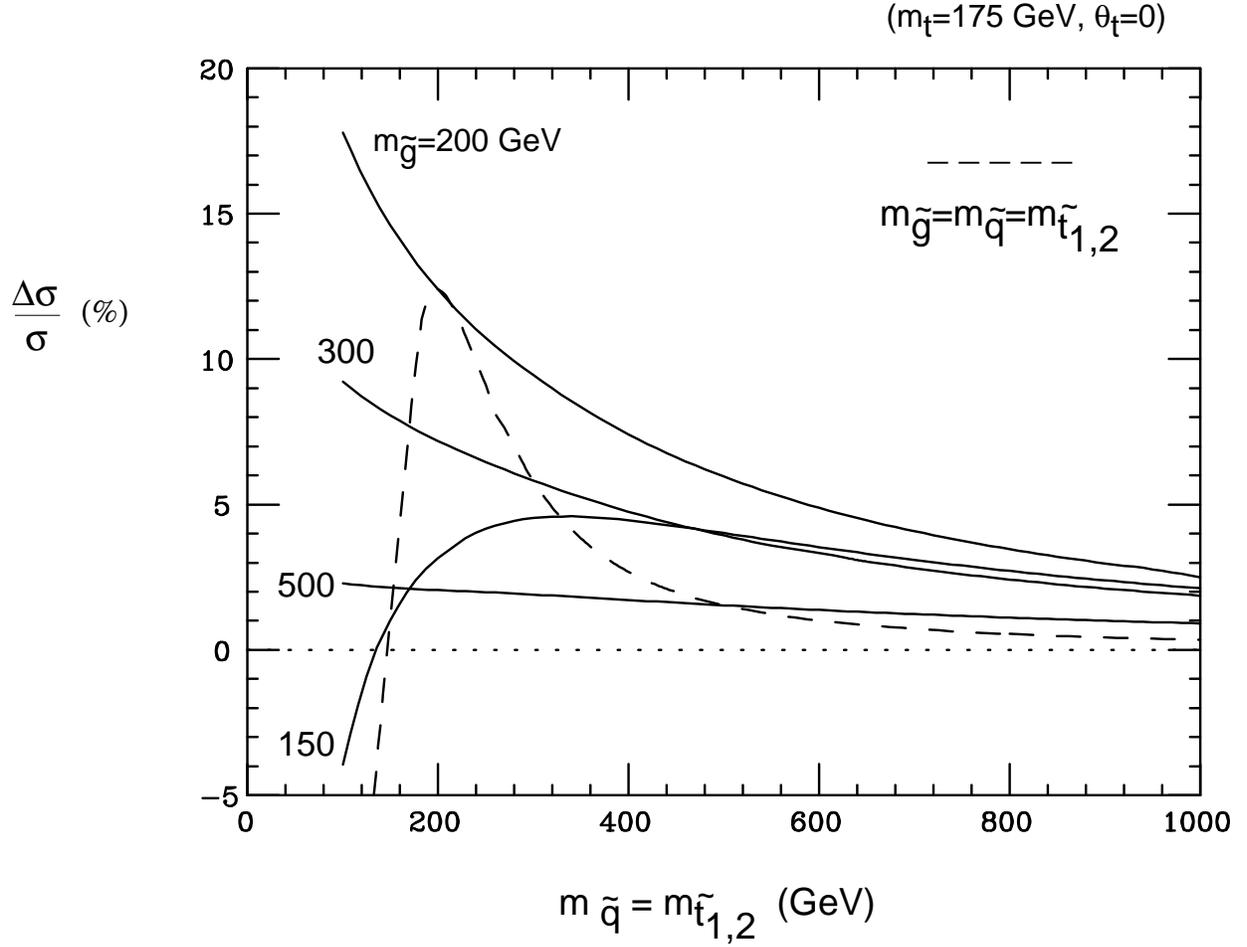}
\caption{The relative (\%) one-loop supersymmetric QCD-like correction to the
top-quark pair-production cross section at the Tevatron as a function of the
universal squark mass, for $m_t=175\,{\rm GeV}$ and the indicated choices of
the gluino mass. The dashed curve represents the $m_{\tilde g}=m_{\tilde q}$
case.}
\label{fig:Unequal}
\end{figure}

As explained above, the QCD-like corrections are not expected to exhibit the dip structure that the electroweak-like corrections possess, because the relation $m_t\approx m_{\tilde g}+m_{\tilde t}$ cannot be satisfied by the
experimentally allowed gluino masses. In Fig.~\ref{fig:Unequal} we show
the QCD-like correction versus the universal squark masses ($m_{\tilde
t_{1,2}}=m_{\tilde q}$) for fixed values of the gluino mass. The curve for
$m_{\tilde g}=150\,{\rm GeV}$ behaves differently from the others because for
sufficiently light squark masses, the relation $m_t\approx m_{\tilde
g}+m_{\tilde q}$ will be satisfied, {\em i.e.}, a dip occurs. This region of
parameter space is disfavored experimentally; it is shown here to make contact
with Ref.~\cite{SUSYQCDe}, with which we agree qualitatively. From
Fig.~\ref{fig:Unequal} we see that shifts as large as $\sim20\%$ may occur.
We also observe that the corrections go to zero when the supersymmetric
particle masses get large, {\em i.e.}, the expected decoupling effect. This
figure also contains a curve (the dashed line) where all sparticle masses are
taken to be the same ($m_{\tilde g}=m_{\tilde t_{1,2}}=m_{\tilde q}$), which is
seen to intercept the other curves at the appropriate places, and to
decouple rather quickly.

The significant QCD-like shifts observed in Fig.~\ref{fig:Unequal}, especially
for $m_{\tilde g}=200\,{\rm GeV}$, may be understood in terms of the large
color factor in the vertex correction diagrams that couple the gluon to two
gluinos, both in the incoming (Fig.~\ref{fig:QCDdiagrams}(c)) and outgoing
(Fig.~\ref{fig:QCDdiagrams}(g)) parts of the diagram. A relevant role may also
be played by a dynamical enhancement of the vertex diagrams that occurs for
$\sqrt{\widehat s}=2m_{\tilde g},2m_{\tilde q}$. As $\sqrt{\widehat s}$ is
integrated over the parton distribution functions, starting at 
$\sqrt{\widehat s}=2m_t$ and peaking at $\sqrt{\widehat s}\approx 400\,{\rm GeV}$, the typical cusp is smeared out but it appears visible as the peak in
the dashed curve in Fig.~\ref{fig:Unequal} (obtained for $m_{\tilde q}=m_{\tilde g}$).

\begin{figure}[p]
\vspace{5in}
\includegraphics{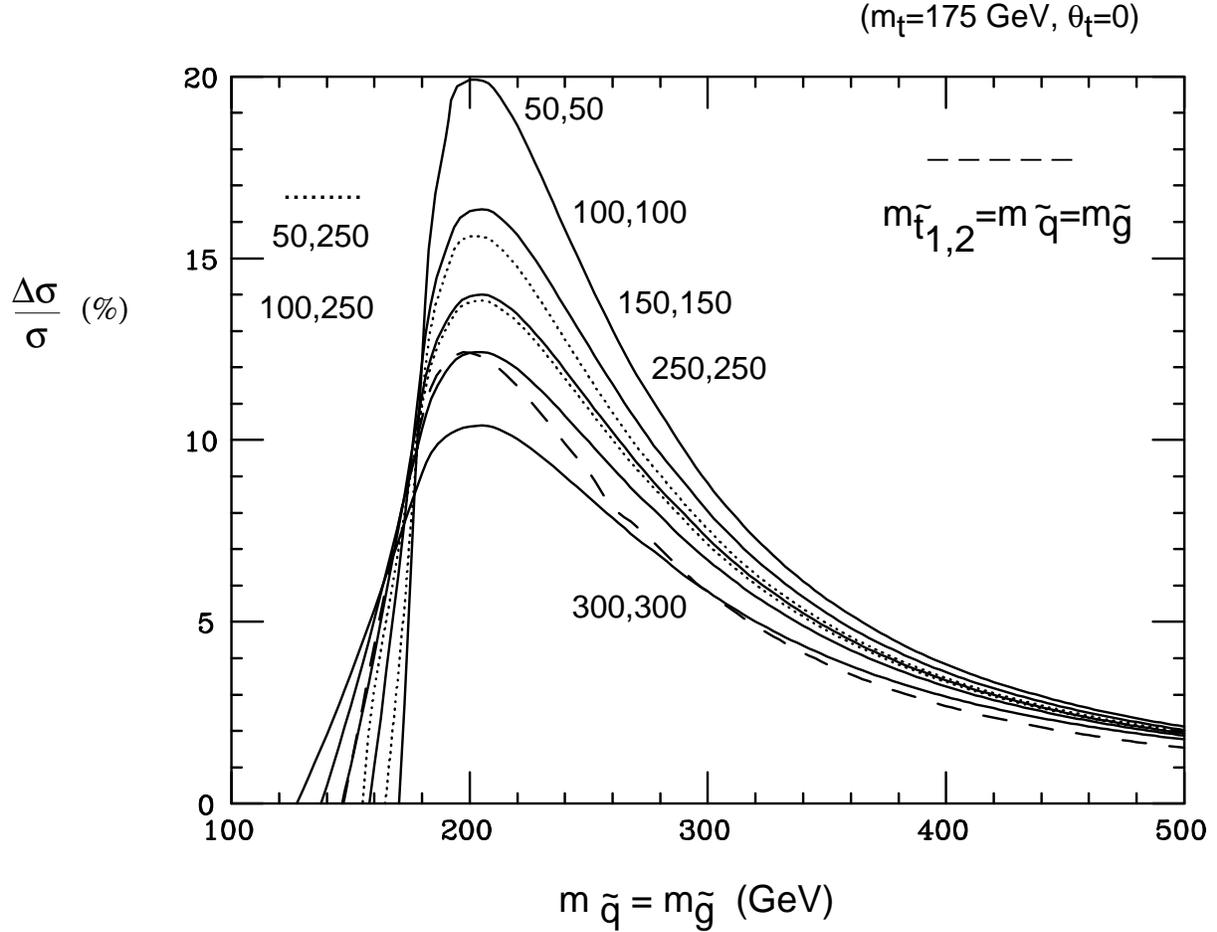}
\caption{The relative (\%) one-loop supersymmetric QCD-like correction to the
top-quark pair-production cross section at the Tevatron as a function of the
squark or gluino mass, for $m_t=175\,{\rm GeV}$ and for the indicated choices
of top-squark masses ($m_{\tilde t_1},m_{\tilde t_2}$). The
dotted curves highlight the effect of non-degenerate top-squark masses. The
dashed curve represents the fully degenerate $m_{\tilde t_{1,2}}=m_{\tilde
g}=m_{\tilde q}$ case.}
\label{fig:Equal}
\end{figure}

In order to explore the effects of lighter top-squark masses, we concentrate
on the $m_{\tilde q}=m_{\tilde g}$ case\footnote{The $m_{\tilde q}\approx
m_{\tilde g}$ relation occurs naturally in supergravity theories.} (the dashed
line in Fig.~\ref{fig:Unequal}) and in Fig.~\ref{fig:Equal} we plot the QCD-like
corrections for representative choices of $(m_{\tilde t_1},m_{\tilde t_2})$.
For reference, the all-equal-masses case is shown as a dashed line (as in
Fig.~\ref{fig:Unequal}). We can see that light top-squark masses enhance the
corrections, especially in the region $m_{\tilde q}=
m_{\tilde g}\approx(200-250)\,{\rm GeV}$. This enhancement occurs (although to
a lesser extent) even if only one of the top-squarks is light. We have also
explored the effect of top-squark mixing, which is non-vanishing only if
$m_{\tilde t_1}\not=m_{\tilde t_2}$. We find this effect to be rather small in
the case of the QCD-like corrections amounting, for example, to a decrease in
the peak value of the (50,250) curve in Fig.~\ref{fig:Equal} by $15\%$ for
maximal mixing.

\section{Expectations in specific models}
\label{sec:models}
The results presented in the previous section should represent the largest
one-loop supersymmetric shifts (due to vertex and external leg corrections)
to be expected in the top-quark cross section. However, in specific regions of MSSM parameter space or specific supergravity models, the shifts are likely to be much smaller than the largest possible ones, as the conditions for enhancement may not be satisfied: large electroweak-like shifts require a higgsino-like neutralino, light top-squarks, $\tan\beta\approx1$, and a specific relation between their masses ({\em i.e.}, $m_\chi+m_{\tilde t}\approx m_t$); large QCD-like shifts require $m_{\tilde q},m_{\tilde g}<250\,{\rm GeV}$ and light top-squarks. Interestingly enough, various scenarios ({\em i.e.}, selected regions of MSSM parameter space) recently proposed to possibly explain some experimental measurements that appear to deviate from Standard Model expectations, fall {\em precisely} in the class of models that may lead to one-loop enhancements of the top-quark cross section.

The discrepancy between the LEP measured value of $R_b$ and its prediction in
the Standard Model may be alleviated by supersymmetric loop corrections to
the $Zb\bar b$ vertex that involve charginos and top-squarks \cite{Rbsusy}.
Moreover, $\tan\beta$ should be close to 1, the charginos should be
higgsino-like (and correspondingly the neutralinos too), and the top-squarks
should be right-handed. Both should be as light as LEP~1.5 searches allow
\cite{LEP15}. Since in this scenario there are no restrictions on the squark or
gluino masses, let us concentrate on the electroweak-like corrections, that will be
significantly enhanced (negatively) in this case if the relation
$m_\chi+m_{\tilde t}\approx m_t$ happens to be satisfied. This may indeed occur
for $m_{\tilde t_2}$ (but not for the desired values of $m_{\tilde t_1}$ and
$m_\chi$, both below $M_W$).

In analogy with the so-called light-gluino window, it has been remarked
that there is a light-higgsino window \cite{Feng}, where the Higgsino mixing
parameter and the SU(2) gaugino mass are very small ($\mu,M_2\approx0$)
and $\tan\beta\approx1$. In this scenario there are three neutralinos with
significant higgsino admixtures, two with masses close to $M_Z$ and a very
light one. Moreover, a light top-squark is also desired to enhance
$R^{\rm susy}_b$. Therefore, if $m_t\approx m_\chi+m_{\tilde t}\approx
M_Z+m_{\tilde t}$ or equivalently
$m_{\tilde t}\approx m_t-M_Z\approx85\,{\rm GeV}$, then the top-quark cross
section will be shifted to lower values by a significant amount.

The last scenario we address has been advanced in Refs.~\cite{Kane,KaneMrenna}
as a possible explanation for the one much-publicized event at CDF, consisting
of $ee\gamma\gamma$ plus missing energy. This event has been ascribed to
selectron pair-production, with decay into electron and second-to-lightest
neutralino ($\chi^0_2$), and further radiative decay of the neutralino into the
LSP ($\chi^0_2\to\chi^0_1+\gamma$). The missing energy is carried away by the
pair of lightest neutralinos produced. The dominance of the (one-loop)
radiative decay $\chi^0_2\to\chi^0_1+\gamma$ over the more traditional ones
($\chi^0_2\to\chi^0_1 f\bar f$) provides the most important constraint on the
parameter space, requiring a higgsino-like $\chi^0_1$ and a photino-like
$\chi^0_2$. This is achieved by setting the SU(2) and U(1) gaugino masses equal
at the electroweak scale ($M_1=M_2$ and $\tan\beta\approx1$), precluding the
gaugino mass unification in GUTs. Furthermore, the kinematics of the event
appear to require: $m_{\chi^0_1}\approx(30-55)\,{\rm GeV}$ and
$m_{\chi^0_2}\approx m_{\chi^0_1}+30\,{\rm GeV}$.
To make contact with the supersymmetric enhancement of $R_b$, it is also
assumed that the top-squark is light ($m_{\tilde t_1}\approx (45-60)\,{\rm
GeV}$). In this case, however, the top-quark would have enhanced decays to
$\tilde t+\chi^0_1$, thus diluting the observed top-quark sample. To undo this
effect, it has been further proposed \cite{KaneMrenna} that squarks and gluinos
should be as light as experimentally allowed ($m_{\tilde
g}\approx(210-235)\,{\rm GeV}$, $m_{\tilde q}\approx(220-250)\,{\rm GeV}$) such
that $\tilde g\to t\tilde t$ decays add to the top-quark sample significantly.
(This mechanism was originally proposed in
Ref.~\cite{Kon}.) Given these values of the squark and gluino masses, from
Fig.~\ref{fig:Equal} one can see that we generally expect $\approx+15\%$
supersymmetric QCD-like corrections to the top-quark cross section. (There may
also be significant electroweak-like corrections if $m_{\tilde t_2}\approx
m_t-m_{\chi^0_1}\approx(105-145)\,{\rm GeV}$.)

Going beyond the specific models discussed above, the light top-squarks that
enhance both electroweak-like and QCD-like corrections will likely decrease the canonical top-quark branching ratio into $bW$ because of the availability of the supersymmetric decay channels $t\to\tilde t+\chi$. Indeed, the top-quark Yukawa coupling in Eq.~(\ref{eq:Yukawa}) entails an enhanced
coupling between the top-quark, a higgsino-like neutralino, and a right-handed
top-squark. This situation is favored by the enhanced corrections discussed
above, and therefore enhance the exotic decays of the top quark. At present
there are no real experimental limits on $B(t\to bW)$, only on $B(t\to
bW)/B(t\to qW)$ \cite{Incandela}. The only limits on  $B(t\to bW)$ have been
obtained by correlating the top-quark mass and cross section measurements with
the Standard Model cross section, implying $B(t\to {\rm other})<25\%$
\cite{WLN}. If the electroweak-like (negative) correction occurs when the exotic channel ($t\to \tilde t\chi$) is kinematically allowed, the yield of $bWbW$ events will be decreased by the two effects, implying an effective $B(t\to bW)$ ratio as small as $(0.65)(0.5)\approx0.3$, assuming a $-35\%$ shift in the cross section and $B(t\to\tilde t\chi)=1/2$. However, we note that the enhancements to the electroweak-like corrections occur right {\em below} the threshold for top-quark decay into top-squark and neutralino (see
Figs.~\ref{fig:50100},\ref{fig:75250}), and therefore the exotic decay channel
is effectively closed.

Recent complementary studies of top-quark properties in supersymmetric theories
include supersymmetric one-loop corrections to the $t\to\tilde t\chi$ exotic
decay channel \cite{DHJ}, supersymmetric three-body decays of the top quark
\cite{GS}, and one-loop supersymmetric corrections to the top-quark width
\cite{Sola}.

\section{Conclusions}
\label{sec:conclusions}
The experimental study of the top quark has just begun in earnest, with its
mass and cross section having been measured to some precision. The Main
Injector upgrade of the Tevatron should essentially provide a top-quark
factory, where observations will be confronted with theoretical expectations
for cross sections and branching ratios. We have shown that these precise
measurements may indeed point to deviations from the Standard Model, as may be
expected in supersymmetric theories. These deviations may be quite sizeable and
therefore easy to detect, especially in scenarios with rather light sparticles
that have been proposed to explain various apparent experimental anomalies. In
the longer term, the presence of top quarks at the LHC will constitute one of
largest backgrounds in new physics searches. Therefore, it will be essential
to have a very good understanding of top-quark physics beforehand, so that
these backgrounds may be subtracted off appropriately.

\section*{Acknowledgments}
We would like to thank Jin Min Yang for useful discussions. R.R. would also
like to thank Toby Falk, Craig Pryor and David Robertson for useful discussions. The work of J.K. and R.R. has been supported by the World Laboratory. The work of J.~L. has been supported in part by DOE grant DE-FG05-93-ER-40717. The work of D.V.N. has been supported in part by DOE grant DE-FG05-91-ER-40633.

\section*{Appendix}
In this Appendix, we present analytical expressions for the one-loop supersymmetric electroweak-like vertex and external leg corrections to top-quark production by including all components of the neutralino (viz., the higgsino, the photino and the zino) and of the chargino (viz., the higgsino and the wino), as well as top-squark and bottom-squark mixing. Though the contributions are enhanced for higgsino-like neutralinos and charginos, in a 
detailed numerical calculation one would also need to include the contributions from gaugino-like neutralinos and charginos. We again ignore corrections arising from box diagrams.

The invariant amplitude for top-quark production via $q\bar q$ annihilation
can be written as
\begin{equation}
M=M_0+ \delta M
\label{eq:A1}
\end{equation}
where $M_0$ is the tree level amplitude and $\delta M$ is the first-order
electroweak-like correction.  $M_0$ and $\delta M$ are given by
\begin{eqnarray}
i M_0&=&\bar v(p_2)(-i g_s T^A\gamma^\nu) u(p_1) 
\,{-i g_{\nu\mu}\over\widehat s}\,
\bar u(p_3) (-i g_s T^A\gamma^\mu) v(p_4)
\label{eq:A2}\\
i \delta M&=&\bar v(p_2) (-i g_s T^A\gamma^\nu) u(p_1) 
\,{-i g_{\nu\mu}\over\widehat s}\,\bar u(p_3) \Lambda^\mu v(p_4)
\label{eq:A3}
\end{eqnarray}
where $p_1, p_2, p_3$ and $p_4$ are the incoming quark and anti-quark and
outgoing top-quark and anti-top-quark momenta respectively.
$\Lambda^\mu$ can be written as
\begin{equation}
\Lambda^\mu=-i g_s T^A \bigl(
F_1 \gamma^\mu +F_2 \gamma^\mu\gamma_5 + F_3 k^\mu
+ F_4 k^\mu\gamma_5 + F_5 i k_\nu\sigma^{\mu\nu} + 
F_6 i k_\nu\sigma^{\mu\nu}\gamma_5
\bigr)
\label{eq:A4}
\end{equation}
where $k^\mu=p_1^\mu+p_2^\mu=p_3^\mu+p_4^\mu$ and 
$\sigma^{\mu\nu}={i\over 2}[\gamma^\mu,\gamma^\nu]$.  The form
factors $F_{1,..,5}$ encode the loop functions and depend on the various masses in the theory. The supersymmetric electroweak-like vertex and external leg corrections to the tree-level parton cross section are give by
\begin{equation}
\Delta\widehat\sigma^{\rm EW}={8\pi\alpha^2_s\over9\widehat s^2}\,\beta_t\,
\left[{\textstyle{2\over3}}(\widehat s+2m^2_t)(F_1^n+F_1^c)
+2(F_5^n+F_5^c) m_t\widehat s\right]\ .
\label{eq:AEWloop}
\end{equation}
We see that the corrections depend only on $F_1$ and $F_5$.  Below we present
the complete form factors that appear in these equations. The tree-level parton cross section is given in Eq.~(\ref{eq:tree}). 

The Feynman rules used to obtain the following form factors are given in
Ref.~\cite{HK}.  For the electroweak-like corrections involving
neutralinos and top-squarks we get
\begin{eqnarray}
F_1^n&=&{1\over 4\pi^2}\sum_{j=1}^4\biggl\{ 
\bigl( |A_n\cos\theta_t + B_n \sin\theta_t|^2 
+ |C_n\cos\theta_t -A_n^* \sin\theta_t|^2 \bigr)\nonumber\\
&&\qquad\qquad\times\left[ c_{24}+m^2_t (c_{11}+c_{21})+{\textstyle{1\over2}}B_1
+m^2_t B'_1\right]^{(\chi^0_j,\tilde t_1,\tilde t_1)}\nonumber\\
&&-{\rm Re}\,\bigl[ (A_n \cos\theta_t + B_n \sin\theta_t)
(C_n^*\cos\theta_t-A_n\sin\theta_t)\bigr]
\, 2 m_t\, m_{\chi^0_j}
\,(c_0+c_{11}+B'_0)^{(\chi^0_j,\tilde t_1,\tilde t_1)}\biggr\}\nonumber\\
&&+ (\tilde t_1\leftrightarrow \tilde t_2,
\,\cos\theta_t\leftrightarrow-\sin\theta_t, 
\,\sin\theta_t\leftrightarrow\cos\theta_t)\ ;
\label{eq:A5}\\
F_5^n&=&{1\over4\pi^2}\sum_{j=1}^4\biggl\{ 
\bigl( |A_n\cos\theta_t + B_n \sin\theta_t|^2 
+ |C_n\cos\theta_t -A_n^* \sin\theta_t|^2 \bigr)
\left[-{\textstyle{1\over2}}m_t(c_{11}+c_{21})\right]
^{(\chi^0_j,\tilde t_1,\tilde t_1)}\nonumber\\
&&+{\rm Re}\,\bigl[ (A_n \cos\theta_t + B_n \sin\theta_t)
(C_n^*\cos\theta_t-A_n\sin\theta_t)\bigr]
\,m_{\chi^0_j}\,
(c_0+c_{11})^{(\chi^0_j,\tilde t_1,\tilde t_1)}\biggr\}\nonumber\\
&&+ (\tilde t_1\leftrightarrow \tilde t_2,
\,\cos\theta_t\leftrightarrow-\sin\theta_t, 
\,\sin\theta_t\leftrightarrow\cos\theta_t)\ ,
\label{eq:A6}
\end{eqnarray}
where (in the notation of Ref.~\cite{HK})
\begin{eqnarray}
A_n&=&-{i\over 2}{g m_t\over\sqrt 2 M_W\sin\beta}N_{j4}^*=
-{i\over2} \lambda_t N_{j4}^*
\label{eq:A5a}\\
B_n&=&{i\over\sqrt 2} \left[{2\over3}e N_{j1}^{\prime *}
-{2\sin^2\theta_W\over3\cos\theta_W}g N_{j2}^{\prime *}\right]
\label{eq:A5b}\\
C_n&=&-{i\over\sqrt 2} 
\left[{2\over3}e N_{j1}^\prime-{2\sin^2\theta_W\over3\cos\theta_W}g
N_{j2}^\prime +{1\over2\cos\theta_W} g N_{j2}^\prime\right]
\label{eq:A5c}
\end{eqnarray}
Recall also that $\tilde
t_1=\cos\theta_t\,\tilde t_L+\sin\theta_t\,\tilde t_R$ and $\tilde
t_2=-\sin\theta_t\,\tilde t_L+\cos\theta_t\,\tilde t_R$.
The various $B$ and $c$ functions in the above expressions are the well
documented Passarino-Veltman functions \cite{PV} (adapted to our metric where
$p^2_i=m^2_i$); the $B$ functions depend on 
$(m_t,m_{\chi^0_j},m_{\tilde t_{1,2}})$ whereas the $c$ functions depend on $(-p_3,p_3+p_4,m_{\chi^0_j}, m_{\tilde t_{1,2}},m_{\tilde t_{1,2}})$ [as reminded by the superscripts in Eqs.~(\ref{eq:A5},\ref{eq:A6})]. We note that
in the expressions for $F^n_{1,5}$ given in the main text in Eqs.~(\ref{eq:F1n},\ref{eq:F5n}), which are specific to the case 
$N_{j1}^\prime=N_{j2}^\prime=0$, we have extracted an overall factor of
$(\lambda_t/4\pi)^2$.

The form factors for the electroweak-like corrections due to loops involving
charginos and bottom-squarks are given by 
\begin{eqnarray}
F_1^c&=&{1\over 4\pi^2}\sum_{j=1}^2\biggl\{ 
\bigl( |A_c\cos\theta_b| ^2 
+ |B_c\cos\theta_b +C_c \sin\theta_b|^2 \bigr)\nonumber\\
&&\qquad\qquad\times\left[ c_{24}+m^2_t (c_{11}+c_{21})+{\textstyle{1\over2}}B_1
+m^2_t B'_1\right]^{(\chi^\pm_j,\tilde b_1,\tilde b_1)}\nonumber\\
&&-{\rm Re}\,\bigl[ A_c \cos\theta_b 
(B_c^*\cos\theta_b+C_c^*\sin\theta_b)\bigr]
\, 2 m_t\, m_{\chi^\pm_j}
\,(c_0+c_{11}+B'_0)^{(\chi^\pm_j,\tilde b_1,\tilde b_1)}\biggr\}\nonumber\\
&&+ (\tilde b_1\leftrightarrow \tilde b_2,
\,\cos\theta_b\leftrightarrow-\sin\theta_b, 
\,\sin\theta_b\leftrightarrow\cos\theta_b)\ ;
\label{eq:A7}\\
F_5^c&=&{1\over4\pi^2}\sum_{j=1}^2\biggl\{ 
\bigl( |A_c\cos\theta_b |^2 
+ |B_c\cos\theta_b +C_c \sin\theta_b|^2 \bigr)
\left[-{\textstyle{1\over2}}m_t(c_{11}+c_{21})\right]
^{(\chi^\pm_j,\tilde b_1,\tilde b_1)}\nonumber\\
&&+{\rm Re}\,\bigl[ A_c \cos\theta_b
(B_c^*\cos\theta_b+C_c^*\sin\theta_b)\bigr]
\,m_{\chi^\pm_j}\,
(c_0+c_{11})^{(\chi^\pm_j,\tilde b_1,\tilde b_1)}\biggr\}\nonumber\\
&&+ (\tilde b_1\leftrightarrow \tilde b_2,
\,\cos\theta_b\leftrightarrow-\sin\theta_b, 
\,\sin\theta_b\leftrightarrow\cos\theta_b)\ ,
\label{eq:A8}
\end{eqnarray}
where (in the notation of Ref.~\cite{HK})
\begin{eqnarray}
A_c&=&{i\over 2}{g m_t\over\sqrt 2 M_W\sin\beta}V_{j2}^*=
{i\over 2}\lambda_t V_{j2}^*
\label{eq:A8a}\\
B_c&=&-{i\over 2} g U_{j1}
\label{eq:A8b}\\
C_c&=&{i\over 2}{g m_b\over\sqrt 2 M_W\cos\beta}U_{j2}=
{i\over 2}\lambda_b U_{j2}\ .
\label{eq:A8c}
\end{eqnarray}
Also, $\theta_b$ is the bottom-squark mixing angle defined such that $\tilde b_1=\cos\theta_b\,\tilde b_L+\sin\theta_b\,\tilde b_R$ and
$\tilde b_2=-\sin\theta_b\,\tilde b_L+\cos\theta_b\,\tilde b_R$. 
In this case the $B$ functions depend on $(m_t,m_{\chi^\pm_j},m_{\tilde b_{1,2}})$ whereas the $c$ functions depend on $(-p_3,p_3+p_4,m_{\chi^\pm_j},
m_{\tilde b_{1,2}},m_{\tilde b_{1,2}})$. Note also that in the expressions for $F^c_{1,5}$ given in the main text in Eqs.~(\ref{eq:F1c},\ref{eq:F5c}), which are specific to the case  $U_{j1}=U_{j2}=0$, we have extracted an overall factor of $(\lambda_t/4\pi)^2$.

\newpage

\clearpage

\clearpage

\clearpage

\clearpage

\clearpage

\clearpage

\clearpage

\clearpage

\end{document}